\documentclass[final]{svjour3}
\pdfoutput=1
\usepackage{graphicx}
\usepackage{rotating}
\usepackage{mathptmx}

\usepackage{xspace}
\usepackage[sort, numbers,square]{natbib}

\usepackage[colorlinks=true, allcolors=blue]{hyperref}
\usepackage{amsmath,amsfonts,amssymb}
\usepackage{mathrsfs}
\usepackage{bm}

\makeatletter
\journalname{Journal of Low Temperature Physics}

\newcommand{\tc}{$T_{\mbox{\scriptsize c}}$\xspace}
\newcommand{\tbath}{$T_{\mbox{\scriptsize bath}}$\xspace}

\newcommand{\pj}{$P_{\mbox{\scriptsize \;J}}$\xspace}
\newcommand{\rn}{$R_{\mbox{\scriptsize N}}$\xspace}

\begin{document}

\title{Advanced ACTPol TES Device Parameters and Noise Performance in Fielded Arrays}

\author{Kevin T. Crowley~\textsuperscript{1} \and Jason E. Austermann~\textsuperscript{3} \and Steve K. Choi~\textsuperscript{1} \and Shannon M. Duff~\textsuperscript{3} \and Patricio A. Gallardo~\textsuperscript{2} \and Shuay-Pwu Patty Ho~\textsuperscript{1} 
	\and Johannes Hubmayr~\textsuperscript{3} \and Brian J. Koopman~\textsuperscript{2} \and Federico Nati~\textsuperscript{6} \and Michael D. Niemack~\textsuperscript{2} \and Maria Salatino~\textsuperscript{1} \and 
	Sara M. Simon~\textsuperscript{4} \and Suzanne T. Staggs~\textsuperscript{1} \and Jason R. Stevens~\textsuperscript{2} \and Joel N. Ullom~\textsuperscript{3} \and Eve M. Vavagiakis~\textsuperscript{2} \and Edward J. Wollack~\textsuperscript{5}}

\authorrunning{Kevin T. Crowley et al.}

\institute{ 	1 Department of Physics, Princeton University, Princeton, NJ 08540, USA, \email{ktc2@princeton.edu}\\ 
	      	2 Department of Physics, Cornell University, Ithaca, NY 14853, USA\\
		3 Quantum Sensors Group, NIST, Boulder, CO 80306, USA\\
		4 Department of Physics, University of Michigan, Ann Arbor, MI 48103, USA\\
		5 NASA Goddard Space Flight Center, Greenbelt, MD 20771, USA\\
		6 Department of Physics and Astronomy, University of Pennsylvania, Philadelphia, PA 19104, USA
}

\maketitle
\begin{abstract}

The Advanced ACTPol (AdvACT) upgrade to the Atacama Cosmology Telescope (ACT) features arrays of aluminum manganese transition-edge sensors (TESes) optimized for ground-based observations of the Cosmic Microwave Background (CMB). Array testing shows highly responsive detectors with anticipated in-band noise performance under optical loading. We report on TES parameters measured with impedance data taken on a subset of TESes. We then compare modeled noise spectral densities to measurements. We find excess noise at frequencies around 100 Hz, nearly outside of the signal band of CMB measurements. In addition, we describe full-array noise measurements in the laboratory and in the field for two new AdvACT mid-frequency arrays, sensitive at bands centered on 90 and 150 GHz, and data for the high-frequency array (150/230 GHz) as deployed.

\keywords{Cosmic Microwave Background, transition-edge sensor, detector modeling, noise performance, TES parameters}

\end{abstract}

\section{Introduction}
\label{Intro}

The AdvACT project comprises four new microwave detector arrays for the focal plane of ACT, a six-meter off-axis Gregorian telescope which observes from the Atacama Desert in northern Chile at a high-altitude (5190 m) site \cite{Thornton}. The science goals of AdvACT include constraining neutrino properties, using Sunyaev-Z'eldovich effects to probe structure, and pursuing primordial B-mode polarization signals. These goals will be met by surveying large sky areas, enabling cross-correlation studies with surveys at different wavelengths \cite{deBernardis:2016}. To constrain large angular-scale polarization, AdvACT will use ambient-temperature continuously-rotating half-wave plates to modulate incoming polarization signals. Achieving these goals requires highly sensitive detector arrays and strict control over instrumental systematics to ensure their effects are subdominant to statistical uncertainties.\\
\indent  The four AdvACT arrays will observe in five total frequency bands, with bandcenters spanning from 27 GHz to 230 GHz \cite{Henderson:2016}. Each array consists of pixels featuring on-wafer orthomode transducers (OMT) followed by filters and hybrid tees within the pixel. Filtered signals terminate in lossy Au meanders, where the dissipated heat is measured by TES bolometers. Each pixel features two pairs of TESes, each pair being sensitive to orthogonal polarizations in two distinct bandpasses. The OMTs are coupled to the telescope via arrays of spline-profiled feedhorns \cite{Simon:2016}.\\
\indent In this proceeding, we focus on laboratory studies of the two mid-frequency (MF) arrays, hereafter MF1 and MF2 in chronological numbering. We seek to confirm that the TES device behavior accords with what we have designated the ``simple" electrothermal model. In Section \ref{devices}, we describe the main features of this model. The experimental configurations for our tests are addressed in Section \ref{experiment}, and laboratory measurements are discussed in Section \ref{lab_results}. We finally discuss performance of the arrays during observations in Section \ref{field}.
\section{TES Models}
\label{devices}

In this section, we briefly review the main features of the simplest TES electrothermal models to be used in this work \cite{IrwinHilton}. A heat capacity $C$ is coupled to a thermal bath (at temperature \tbath) via a thermal link with zero heat capacity and thermal conductance $G$. With regard to electrical properties, we view the TES as a temperature- and current-sensitive resistor. The maximum resistance, \rn, occurs for a normal-state TES. This parameter, and the TES critical temperature \tc, are important for ensuring device operability and performance.\\
\indent When the TES operates, heat dissipated by microwave signals $P_{\mbox{\scriptsize opt}}$ and Joule heating of the TES \pj is balanced by $P_{\mbox{\scriptsize th}}$, the thermal power flowing to the bath. We index steady-state conditions using the parameters $R_0$, the steady-state device resistance, and $P_J$. In the steady state, the TES sensitivities to changes in temperature and current are expressed as $\alpha = \frac{dlnR}{dlnT}$ and $\beta = \frac{dlnR}{dlnI}$. We will seek to constrain these parameters based on measurements of the TES impedance. The impedance as a function of frequency $\omega$ is:
\begin{figure}[h!]
  \centering
  \includegraphics[width=0.66\textwidth]{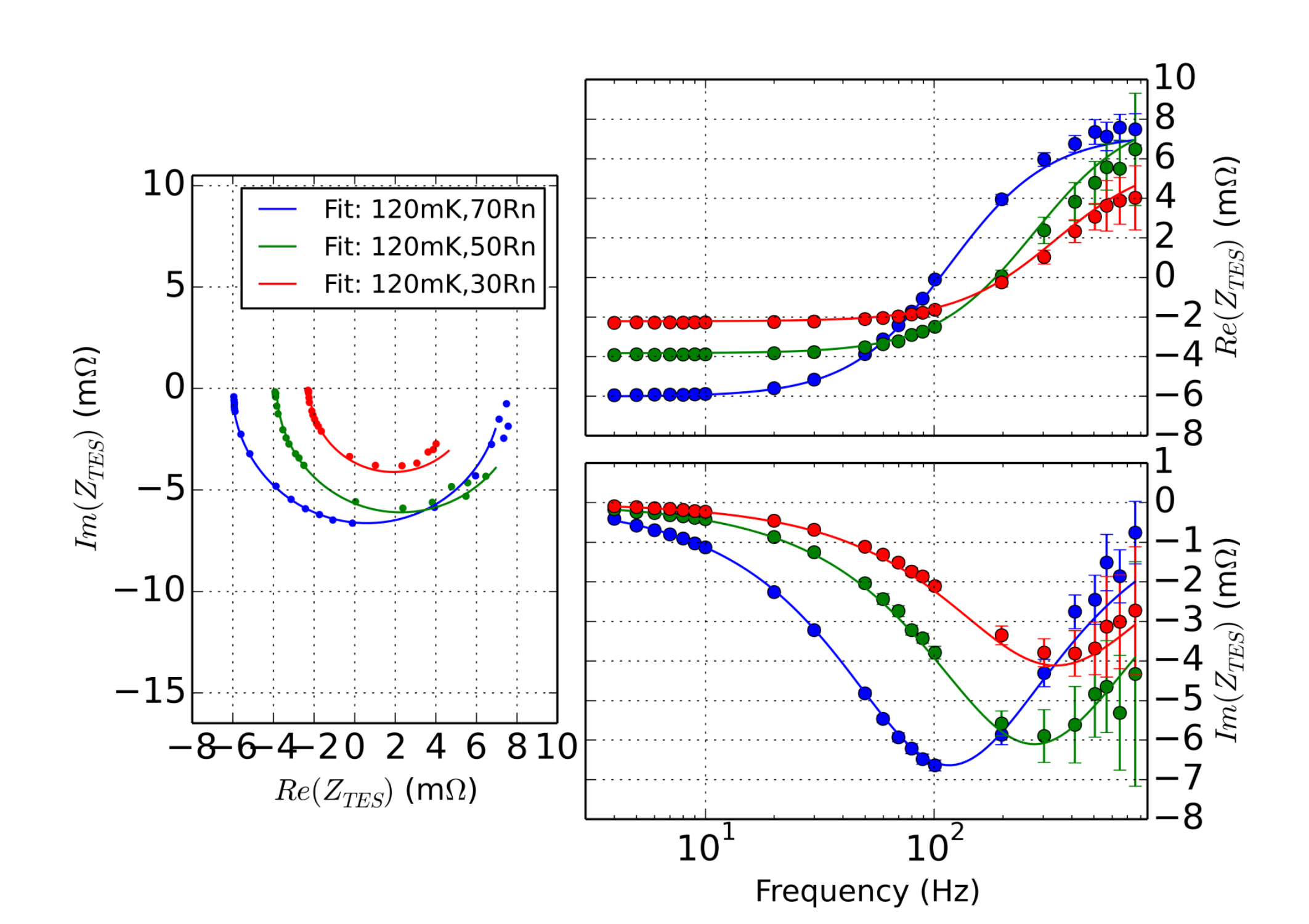}
  \caption{AdvACT impedance data across the transition at 120 mK shown in the complex plane (\textit{left}) and as the real part (\textit{top right}) and imaginary part (\textit{bottom right}) of $Z_{\mbox{\scriptsize TES}}$ versus frequency. Error bars are analytically derived from amplitude and phase errors on the sinusoidal fit to data at each frequency. Solid best-fit curves were produced with seven parameters, resulting in a reduced $\chi^2$ = 88.9/59 = 1.51.}
  \label{imped_example}
\end{figure}
\begin{equation}
\label{impedance}
Z_{\mbox{\scriptsize TES}} = R_0 (1 + \beta) + R_0 (2 + \beta) \frac{\mathscr{L}}{1 - \mathscr{L} + i \omega \tau} ,
\end{equation}
where $\tau = C/G$ is the intrinsic thermal relaxation time and $\mathscr{L}$, the loop gain, parametrizes the constant-current sensitivity of TES resistance to power: $\mathscr{L} = dlnR/dlnP = P_{\mbox{\scriptsize J}} \alpha / G T_{\mbox{\scriptsize c}}$. An example of fitting this equation to an impedance dataset gathered on an AdvACT TES can be seen in Fig. \ref{imped_example}. The semicircular shape of the solid best-fit curves results from Eq. \ref{impedance}.\\
\indent In general, we do not expect the simple thermal model of the TES (i.e. a single lumped-element heat capacity) to hold for all excitation frequencies. In an AdvACT pixel, the TES is part of a larger structure which includes deposited AlMn which is not part of the sensor, and a top layer of PdAu \cite{Duff:2016}. The combined heat capacities of these films determines the effective heat capacity $C$, while the larger structure (the ``island") is isolated from the thermal bath by silicon nitride legs (defining $G$). 
We expect finite-conductance effects between the region of the island thermalized with the sensor and the bulk PdAu. Such effects have been measured in similar devices \cite{George}. High-frequency impedance measurements of separate AdvACT TESes indicate deviations from Eq. \ref{impedance} for frequencies of $\sim$ few kHz, to be discussed in future publications.
\section{Experimental Setup}
\label{experiment}

Below, we describe the hardware and software used to perform TES tests. The TES is voltage-biased within a circuit with a parallel shunt resistor with resistance $R_{\mbox{\scriptsize shunt}}$ and a series inductance $L$. These two bias-circuit components form an anti-aliasing filter. Both laboratory and field measurements of the TES use superconducting quantum interference devices (SQUIDs) to read out TES current signals and multiplex those signals at cryogenic stages \cite{Henderson:2016_2}. The time-division multiplexing of TES signals uses Multi-Channel Electronics (MCE) \cite{Battistelli} to provide SQUID bias voltages and feedback signals, TES bias currents, and multiplexing switching.\\
\indent Lab tests of MF1 and MF2 were performed using an Oxford\footnote{https://www.oxford-instruments.com} Triton 200 dilution refrigerator. A LakeShore\footnote{https://www.lakeshore.com} 370 AC controller was used to maintain stable cryogenic temperatures during data acquisition, with the array temperature tunable from below our calibration scale ($<$ 60 mK) to above 190 mK. During MF1 testing, one-third of the detectors ($\sim$ 600) were illuminated by a blackbody cold load mounted on the 4 K stage of the refrigerator used to perform efficiency measurements \cite{Choi:2017}. Fewer ($\sim$50) detectors were illuminated for MF2 tests, and copper conical horns were used. All of the data discussed here include only unilluminated or ``dark" detectors. In the field, individual AdvACT arrays are loaded into an ``optics tube," of which there are three in the current ACTPol cryostat \cite{Thornton}. This cryostat is cooled by one Cryomech \footnote{http://www.cryomech.com} PT410 pulse tube cooler (PTC), one Cryomech PT407 PTC, and a Janis Research Corporation \footnote{225 Wildwood Ave, Woburn, MA 01801} dilution refrigerator. During observations, the array bath temperature is $\sim$ 100~mK.
\section{Laboratory Results}
\label{lab_results}
\label{lab}

To acquire impedance data, we generate a sine wave using the MCE and add the signal to the constant bias value on the TES bias line. We configure the MCE readout to sample a single detector at 9 kHz rather than 400 Hz. By recording the response to sine waves at many frequencies, we map out the TES transfer function. For any given steady-state condition (hereafter referred to as a ``bias point"), we can calibrate the measured transfer function to account for the TES bias circuit \cite{Lindeman, Zhao}. We measured the calibrated impedance for 10 TESes over both MF arrays at different bath temperatures, thus changing \pj, and at different TES resistances, measured in \%\rn.\\
\begin{figure}[t!]
  \centering
  \includegraphics[width=0.83\textwidth]{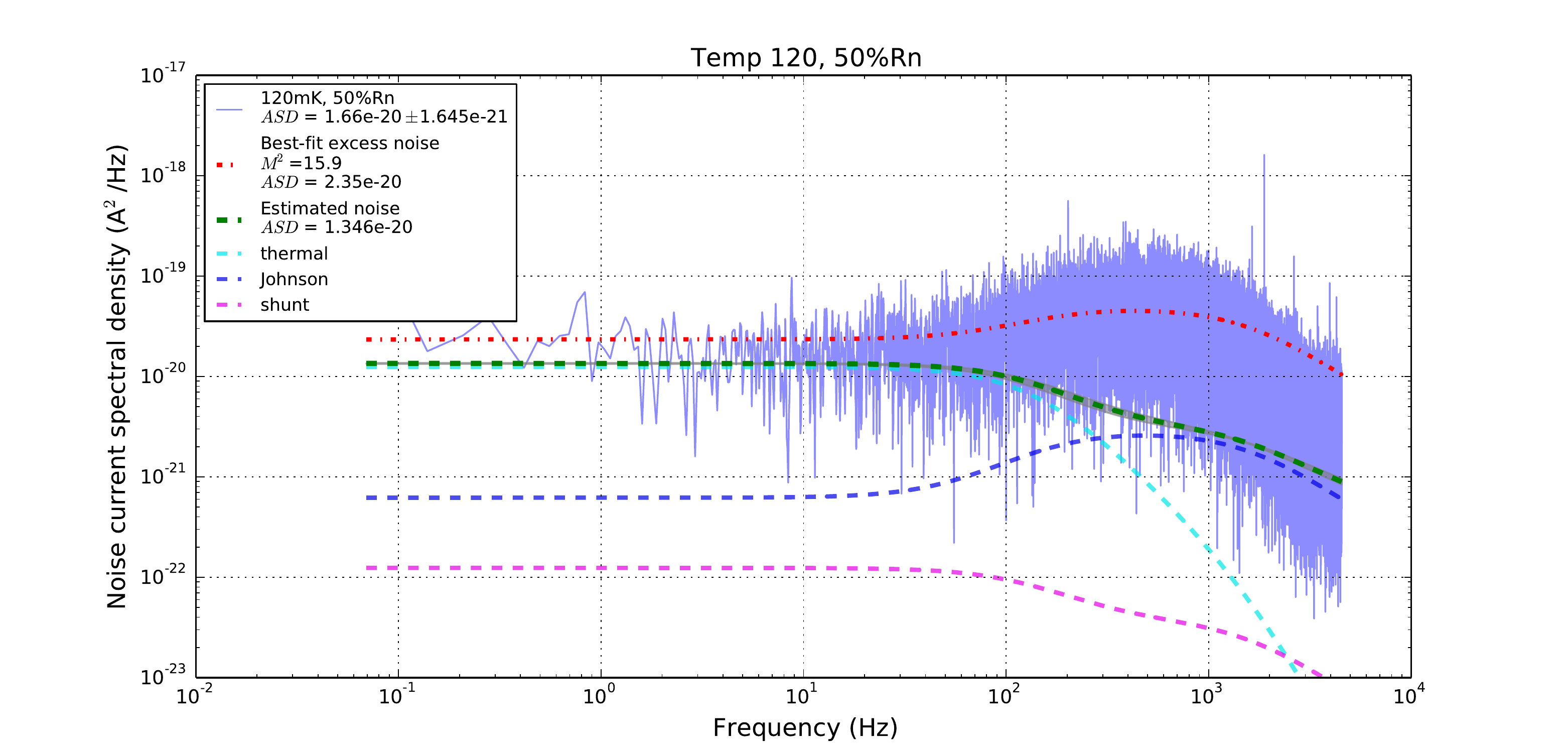}
  \caption{Measured noise current spectral density (blue solid) sampled at 9 kHz for the detector with impedance shown in Fig. \ref{imped_example}. The simple noise model (green dashed) uses impedance best-fit parameters; the gray band is the 68\% spread when drawing 100 realizations of $(\alpha, \beta, C)$ from a multivariate Gaussian defined by the best-fit parameters and covariance. The excess-noise model (red dash-dot) fits a scaling parameter to the TES Johnson noise. Aliasing effects from estimated noise at frequencies above 4.5 kHz are included in both models.}
  \label{noise_ex}
\end{figure}
\begin{table}[b!]
\centering
\caption{MF detector parameters at \%\rn = 50 and \tbath $\in \{$120, 130$\}$ mK for eight detectors. The first column shows medians across detectors, while ``Typical Uncertainties" are estimated statistical errors, averaged across detectors, as percentages of the median parameters. The last column shows differences between the maximum and minimum parameters for the eight detectors. Primary fit parameters are bolded. Here, $S_I$ is the TES power-to-current responsivity, $dI/dP$, at 10 Hz. The quantity $1+M^2$ is discussed in Section \ref{lab}.}
\label{params}
\begin{tabular}{c|c|c|c}
\hline\noalign{\smallskip}
Parameter & Median & Typical Uncertainties (\%) & Range\\
\hline\noalign{\smallskip}
$\bm{C [pJ/K]}$ & 3.8 & 2 & 1.51\\
$\bm{\alpha}$ & 114 & 6 & 63\\
$\bm{\beta}$ & 1.6 & 14 & 0.9\\
$\mathscr{L}$ & 21 & 6 & 16\\
$f_{\mbox{\scriptsize 3dB}}$ [Hz] & 126 & 4 & 69\\
$| S_I |$ (10 Hz) [uA/pW] & 5.0 & -- & 1.5\\
$1 + M^2$ & 8.7 & -- & --\\
\hline
\end{tabular}
\end{table}
\indent To extract parameters, we minimize $\chi^2$ of the datasets at different TES bias points, with error bars analytically combining amplitude and phase errors from sinusoid fits to timestreams at each frequency. We hold $C$ constant across \%\rn datasets for a given bath temperature. Our parameters depend on measured \tc and $G$ from I-V curve datasets discussed in \cite{Choi:2017}. In Tab. \ref{params}, we provide TES parameters for all MF TES datasets with \%\rn = 50 and \tbath $\in \{$120, 130$\}$ mK. These conditions correspond most closely to the bias point we expect for the MF arrays under optical loading. The measured characteristic response frequency, $f_{\mbox{\scriptsize 3dB}}$, which we estimate from impedance and I-V parameters as:
\begin{equation}
f_{\mbox{\scriptsize 3dB}} = \frac{G}{2 \pi C} \left( 1 + \frac{\mathscr{L}}{1 + \beta} \right),
\label{f3db_eff_def}
\end{equation}
has a central value which compares well with values derived from an independent probe in \cite{Choi:2017}. We have made the assumption that we satisfy the condition $R_{\mbox{\scriptsize shunt}} / R_0 \ll 1$, a characteristic of the stiffness of the voltage bias that we assume applies for all operating conditions.\\ 
\indent In fitting the data, we find $\alpha$ and $\beta$ have a high correlation $\sim$ 0.9. Upon investigation, data at frequencies above $\sim$ 400 Hz are downweighted due to higher estimated error, weakening the constraint on $\beta$ from comparing $Re\{ Z_{\mbox{\scriptsize TES}} \}$ at high frequencies to $R_0 (1 + \beta)$. We retain a joint constraint on $\alpha$ and $\beta$ from the semicircle radius $R_Z = - \frac{R_0 (2 + \beta) \mathscr{L}}{2 (1 - \mathscr{L})}$. Constraints on $\alpha$ are further improved by fitting the effective time constant $\tau_{\mbox {\scriptsize I}} = \tau / (1 - \mathscr{L})$.\\
\indent After collecting impedance datasets, we measure detector noise at the 9 kHz sample rate and at the same bias points. We then compare the measured noise current spectral densities (or $ASD$) to the expected noise according to Section 2.6 of \cite{IrwinHilton}. In general, we assume that phonon noise due to $G$ should dominate noise at low frequencies, and that this noise floor rolls off first at the detector $f_{\mbox{\scriptsize 3dB}}$, then at the electrical pole $f_{\mbox{\scriptsize el}} \sim R/(2 \pi L)$.
\begin{figure}[t!]
  \centering
  \includegraphics[width=0.93\textwidth]{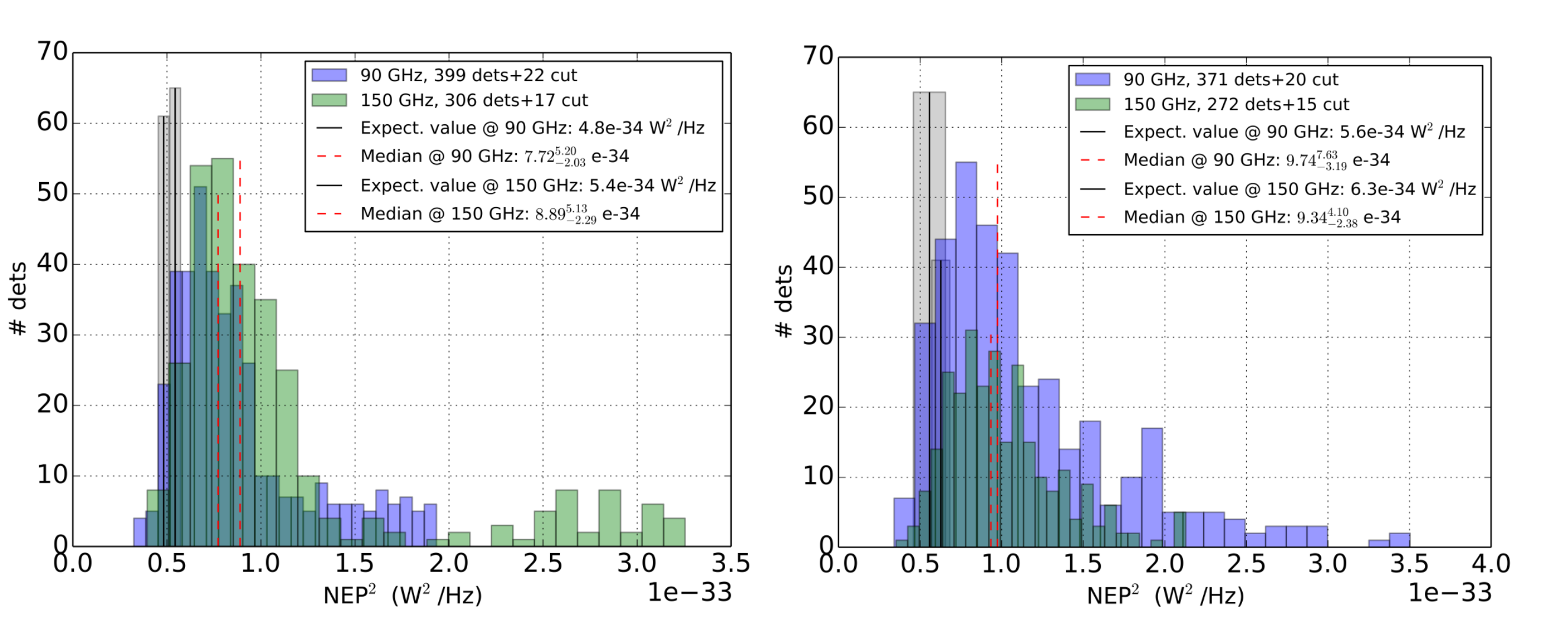}
 \caption{Estimated $NEP^2$ (solid black vertical line) at 10 Hz across the MF1 (left) and MF2 (right) arrays for 400 Hz-sampled lab data at 120 mK \tbath and 50\%\rn with no optical loading. The gray bands define the 1-$\sigma$ spread in expected $NEP^2_{\mbox{\scriptsize ph}}$. The upper 5\% of the data have been excluded from the median estimate (red dashed vertical line) to reject outlier spectra, which generally have an unphysical spectral shape. In the plot, $f_{\mbox{\scriptsize link}}$ is set to 1; when estimated as in \cite{Mather}, the MF arrays have median $f_{\mbox{\scriptsize link}} \sim$  0.7. See text for discussion of this choice.}
  \label{noise_hist}
\end{figure}
For all AdvACT detectors, including those in the high-frequency (HF) array \cite{Crowley:2016}, we observe an excess at frequencies above 100 Hz in dark noise spectra. An example is shown in Fig. \ref{noise_ex}, with data in blue and the expected spectrum in dashed green. To explore this excess, we treat it as due to scaled TES Johnson noise, as in \cite{Jethava:2009}. The model with the best-fit value of the parameter $M^2$, with the Johnson noise scaled by $(1+M^2)$, is also shown. This parameterization makes the minimum value of the fit parameter equal zero and ensures that the overall scaling is positive-definite. We quote a median value in Tab. \ref{params} while noting that this model only provides satisfactory fits for some spectra. The many parameters determining the spectral shape of the Johnson noise make it difficult to diagnose any parameter-dependent effects on the model goodness-of-fit; we will explore this excess in future work.\\
\indent At low frequencies, we find good agreement between model and data for these tests. Explicitly, we take the phonon noise current spectral density to be $ASD_{\mbox{\scriptsize ph}} = 4 k_b G T^2_{\mbox{\scriptsize c}} f_{\mbox{\scriptsize link}} | S_I |^2$ and set the nonequilibrium correction $f_{\mbox{\scriptsize link}}$ = 1 throughout this work. We have done so due to our measurements of $NEP^2$ at different \tbath versus estimated $ f_{\mbox{\scriptsize link}}$ at these temperatures. These data did not show the expected linear trend.\\
\indent For the eight detectors whose parameters went into Tab. \ref{params}, we calculate the average discrepancy as a percentage of the expected model value. We find values at 10 Hz to be  -30\% and -5\%, with and without Johnson noise scaling, respectively.\\
\indent In Fig. \ref{noise_hist}, we show histograms of measured noise-equivalent power, or $NEP^2$, at 10 Hz compared to our expectations. Detectors with unphysical I-V parameters have been cut.
We find that the median measured $NEP^2$ is up to 60\% higher than the median expected $NEP^2$. However, the estimated noise value does not include any aliasing of either SQUID amplifier noise or excess Johnson noise. Note that these estimates include only dark detectors; the discrepancies are less significant once photon noise is included (see Section \ref{field}).\\
\section{Field Performance}
\label{field}
\begin{figure}[b!]
  \centering
  \includegraphics[width=1.02\textwidth]{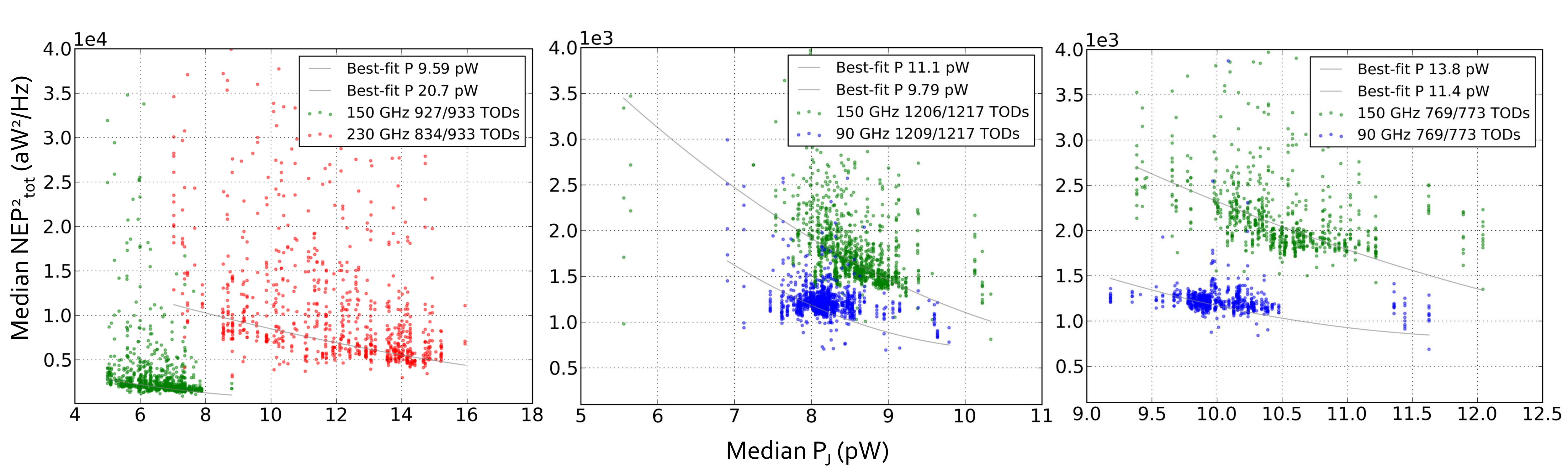}
  \caption{Array median $NEP^2$ at 10 Hz plotted versus median \pj for summer 2017 observations (known as ``TODs") with all three fielded AdvACT arrays (HF, MF1, MF2 from left to right respectively). We take \pj as a proxy for optical loading $P_{\mbox{\scriptsize opt}}$, with low loading at right and high loading at left. Each point corresponds to an AdvACT observation. Gray lines are best-fit lines to the model discussed in the text, with all observations included in the fit. Note the difference in scale between the HF and two MF array plots.}
  \label{field_nep}
\end{figure}
To further study array noise, we investigate estimated $NEP^2$ values of each array in the field during CMB observations. We calibrate detector timestreams
using I-V curve-derived responsivities estimated before observations. Data for detectors with unphysical I-V parameters are removed. The remaining data are filtered with a 3rd-order polynomial before estimating $NEP^2$ at 10 Hz. We then estimate an expected total $NEP^2_{\mbox{\scriptsize tot}}$ combining photon noise as a function of loading power $P_{\mbox{\scriptsize opt}}$ derived from \cite{Lamarre} and measured dark $NEP^2$:
\begin{equation}
NEP^2_{\mbox{\scriptsize tot}} = 2 h \nu_c P_{\mbox{\scriptsize opt}} + 2 \frac{P^2_{\mbox{\scriptsize opt}}}{\Delta \nu} + \mbox{median} ( NEP^2_{\mbox{\scriptsize dark}} ),
\label{NEP_photon}
\end{equation}
for a single-polarization detector, with $\nu_c$ the center frequency of the detector band, $\Delta \nu$ the bandwidth, and the median $NEP^2$ is estimated in each frequency band at bath temperatures for which the dark-detector \pj is closest to \pj measured on the telescope. These temperatures are 120 mK (HF) and 130 mK (MF1/2). 
To fit the field data on $NEP^2_{\mbox{\scriptsize tot}}$, we introduce a parameter $\mathscr{P}$ to define our estimate for $P_{\mbox{\scriptsize opt}}$ in Eq. \ref{NEP_photon}:
\begin{equation}
P_{\mbox{\scriptsize opt}} = \mathscr{P} - \mbox{median} ( P_{\mbox{\scriptsize J,field}} ),
\label{popt_def}
\end{equation}
where the median is over detectors in one frequency band in one array for each $\sim$10 min timestream. The bandcenters $\nu$ are taken to be 97.8 GHz and 147.5 GHz for MF1/2 based on simulated bandpasses as in \cite{Choi:2017}. Bandwidths $\Delta \nu$ are assumed to be 29 GHz (90 GHz) and 40 GHz (150 GHz) from the same simulations. For the HF array, we assume nominal 150 and 230 GHz bandcenters and a bandwidth of 80 GHz for the 230 GHz channel.\\ 
\indent In principle, the fit parameter $\mathscr{P}$ defined by Eq. \ref{popt_def} should equal \pj of the dark detectors in laboratory data at the bath temperature at which we observe, which is approximately 100 mK. In Fig. \ref{field_nep}, we fit Eq. \ref{NEP_photon} to the median $NEP^2$ and \pj values across the three fielded AdvACT arrays. The values of $\mathscr{P}$ indicated in the legend are consistently 2-3 pW lower than the median \pj measured in the laboratory at 100 mK. Known systematics include calibration differences between thermometers in the field and in the laboratory, and uncertainty in resistances in the bias and feedback paths. When comparing the fit to the data, we see acceptable qualitiative agreement. With regard to dispersion in the 230 GHz data, we expect that these detectors, with their large $G$ values, will be the most sensitive to bath temperature fluctuations.\\
\indent We can now determine the ratio between measured laboratory dark $NEP^2$ and the fit to total $NEP^2$ at a nominal value of precipitable water vapor (PWV). Averaging over PWV values between 0.9 and 1.1 mm, we find that ratios of 32\% (HF 150 GHz), 23\% (HF 230 GHz), 31\% (MF1 150 GHz), 38\% (MF1 90 GHz), 26\% (MF2 150), and 42\% (MF2 90). Thus the modest increases in the laboratory dark noise do not have a large effect on field performance.\\
\section{Conclusion}
\label{conclusion}

In this proceeding, we have discussed TES bolometer parameters and noise performance for the three AdvACT arrays which are currently observing the sky in Chile. Overall, we find these AlMn TES bolometers to be responsive and stable with a modest amount of excess noise which, in some cases, can be phenomenologically described with a scaling \mbox{parameter} for the TES Johnson noise. We plan to improve these findings based on more detailed \mbox{electrothermal} models and impedance data with greater spread in frequency. The fielded results indicate that photon noise remains the dominant noise source under typical sky loadings during observations.\\
\begin{acknowledgements}
This work was supported by the U.S. National Science Foundation through award 1440226. The development of multichroic detectors and lenses was supported by NASA grants NNX13AE56G and NNX14AB58G. The work of KTC and BJK was supported by NASA Space Technology Research Fellowship awards.
\end{acknowledgements}

\pagebreak

\end{document}